\documentclass[a4paper,fleqn]{cas-dc}

\usepackage[authoryear,longnamesfirst]{natbib}
\usepackage{cleveref}

\usepackage{bm}
\def\tsc#1{\csdef{#1}{\textsc{\lowercase{#1}}\xspace}}
\tsc{WGM}
\tsc{QE}
\begin{document}
\let\WriteBookmarks\relax
\def\floatpagepagefraction{1}
\def\textpagefraction{.001}

\shorttitle{}    

\shortauthors{}  

\title [mode = title]{Cosmological consequences of a dynamical dark matter in the light of DESI DR2 measurements}  

\tnotemark[1] 

\tnotetext[1]{} 


\author[1]{Abhijith Ajith}[orcid=0009-0009-0213-2406]

\cormark[1]

\fnmark[1]

\ead{abhijith.ajith.421997@gmail.com}

\ead[url]{https://abhijith-78.github.io/}

\credit{}

\affiliation[1]{organization={Indian Institute of Science Education and Research},
            city={Bhopal},
%          citysep={}, % Uncomment if no comma needed between city and postcode
            postcode={462066}, 
            country={India}}

\author[2]{Utkarsh Kumar}[orcid=0000-0002-8158-2078]

% Footnote of the second author
\fnmark[2]

% Email id of the second author
\ead{utkarshkumar.physics@gmail.com}

\ead[url]{}

\credit{}

\affiliation[2]{organization={Department of Physics, University of Ottawa},
            city={Ottawa},
%          citysep={}, % Uncomment if no comma needed between city and postcode
            postcode={ON K1N6N5}, 
            country={Canada}}

\cortext[1]{Corresponding author}

\fntext[1]{}


\begin{abstract}
Recent DESI results exhibit preference for a Null Energy Condition violating dynamical dark energy, with early phantom behaviour. We explore an alternative interpretation in which this preference arises from unconventional dark matter dynamics. We propose a dynamical dark matter (DDM) model, with a non-zero equation of state (EoS) that smoothly interpolates between early time and late time asymptotes across a transition scale factor $a_t$, and confront it against cosmological datasets including CMB, DESI DR2 BAO, SNeIa and growth rate data. We find the early time EoS to be consistent with zero, while the present day value is negative at a significance ranging from $0.42\sigma$ to $1.66\sigma$ depending on the CMB, BAO and SNeIa combinations considered. The strongest preference occurs from the Planck+DESI+DES-Dovekie combination giving the present day EoS to be $-0.045^{+0.020}_{-0.034}$ and $a_t = 0.49^{+0.12}_{-0.23}$ at 68\% CL. This preference for a non-zero, late time DM EoS persists when growth rate data are included and across all three SNeIa compilations considered, while $\Omega_m$ mildly shifts to higher values relative to $\Lambda$CDM. The model also predicts a lower $\sigma_8$ and $S_8$ than $\Lambda$CDM, in better agreement with weak-lensing data, while $H_0$ remains unchanged and in tension with local distance-ladder measurements. The DDM model is preferred over $\Lambda$CDM ($\Delta \chi^2_{\rm MAP} = -10.975$, $\Delta {\rm DIC} = -3.905$ for Planck+DESI+DES-Dovekie) but disfavoured relative to the CPL parameterization of DE ($\Delta \chi^2_{\rm MAP} = 2.975$, $\Delta {\rm DIC} = 5.249$). This preference is consistent among other dataset combinations too. \nocite{*}%% Remove this line from your manuscript.
\end{abstract}

%\begin{graphicalabstrCosmological consequences of a dynamical dark matter in the light of DESI DR2 measurementsact}

%\item DESI DR2 hints of evolving dark energy are reinterpreted as dark matter dynamics.
%\item A time dependent dark matter equation of state avoids NEC-violating dark energy.
%\item Cosmological datasets mildly favour a negative late-time dark matter equation of state.

%\nocite{*}

% Keywords

\maketitle

% Main text
\section{Introduction}\label{}
Cosmological measurements over the past few decades have established the $\Lambda$CDM model as the standard model of cosmology. Observational evidence suggest that the late-time Universe is dominated by a dark sector composed of a dark energy (DE) and dark matter (DM) components which together account for about 95\% of the total energy budget of the Universe. DE is primarily responsible for the late-time acceleration of the Universe, while DM accounts for the formation of large scale structure (LSS) in the Universe. The $\Lambda$CDM model assumes that the DE component is a cosmological constant $\Lambda$ with an equation of state (EoS) parameter $w=-1$ and the DM is cold and pressure-less with $w=0$ (CDM). The $\Lambda$CDM model has been successful in explaining a wide range of cosmological observations, such as the cosmic microwave background (CMB) anisotropies \cite{Planck:2018vyg,Planck:2019nip,Planck:2018lbu, SPT-3G:2021eoc, ACT:2020frw, WMAP:2012nax, Planck:2018nkj,Lemos:2023rdh,Tristram:2007zz}, LSS measurements including the baryon acoustic oscillations (BAO) \cite{BOSS:2016wmc,eBOSS:2020tmo,eBOSS:2020gbb,eBOSS:2020yzd,eBOSS:2020hur,eBOSS:2020lta,DESI:2025fxa,DESI:2024mwx}, and the luminosity distance measurements from Type Ia supernovae (SNeIa) \cite{Pan-STARRS1:2017jku,Scolnic:2021amr,Rubin:2023ovl,Bailey:2022pax,Blondin:2024fpr,DES:2024jxu,NearbySupernovaFactory:2015pcf}.

Despite the tremendous success of the $\Lambda$CDM model, it is severely challenged by the issues related to the $\Lambda$ and CDM components. The consideration of a cosmological constant as the DE component leads to the well known cosmological constant problem, which arises from the discrepancy between the observed value of $\Lambda$ and the theoretical predictions from quantum field theory. Further, with the improvements in measurement precision over the years, multiple discrepancies have appeared between the probes of the early and late Universe \cite{Verde:2019ivm,Hildebrandt:2016iqg,DES:2017myr,Planck:2018vyg,Pedrotti:2024kpn,Vagnozzi:2021tjv,Reeves:2022aoi,Vagnozzi:2019ezj}, which question the underlying nature of the Universe's components. The Hubble tension refers to the 4 to 6$\sigma$ discrepancy between the $H_0$ measurements from Planck ($H_0 = 67.36 \pm 0.54$ km/s/Mpc \cite{Planck:2018vyg}) and the late-time measurements from the SH0ES collaboration ($H_0 = 73.04 \pm 1.04$ km/s/Mpc \cite{Riess:2021jrx}). In addition, the $S_8$ tension identifies the discrepancy between the value  of the matter fluctuation amplitude parameter $S_8$ obtained from the CMB by Planck \cite{Planck:2018vyg}, $S_8 = 0.834 \pm 0.016$, and the directly measured values from KiDS-1000 \cite{Hildebrandt:2016iqg} and DES Y3 \cite{DES:2021wwk}, $S_8= 0.769\pm0.016$, indicating a 2.9$\sigma$ deviation. However, in recent times, with improved calibration, redshift distribution and new survey areas, KiDS measurements \cite{Wright:2025xka} has reported the value of $S_8$ to be $0.81^{+0.016}_{-0.021}$. This value is in close agreement ($0.73\sigma$) with the results from Planck Legacy CMB observations. Moreover, such disagreements in the context of the $\Lambda$CDM have motivated the exploration of alternative DE models, such as quintessence, k-essence, quintom, and modified gravity theories \cite{PhysRevD.37.3406,Chiba:1997ej,PhysRevD.62.023511,PhysRevD.63.103510,Sotiriou:2008rp,Manoharan:2022qll,Rinaldi:2014yta,Hu:2007nk,Casalino:2018tcd,Mukhopadhyay:2019cai,Ruchika:2020avj,Odintsov:2020zct,Ong:2022wrs,Luciano:2023wtx,Tyagi:2025zov,Akrami:2025zlb,Santos:2025wiv,Colgain:2025nzf,Artymowski:2020zwy,Artymowski:2021fkw,Ben-Dayan:2023rgt,Ben-Dayan:2023htq,DiValentino:2020kpf,Bella:2026zuk,DiValentino:2019ffd,Vagnozzi:2018jhn,Li:2026asg,Du:2026qtq,RoyChoudhury:2025iis,Chakraborty:2025syu,DOnofrio:2025cuk,Ren:2026jyw,Koutroulis:2026gjr}.

On the other hand CDM explains the formation of structures in the Universe at the scales much larger than 1 Mpc. However, it faces challenges at smaller scales \cite{Copi:2013cya, Copi:2006tu,Bullock:2017xww,Ben-Dayan:2024uvx}, such as the core-cusp problem \cite{deVega:2013jfy}, the missing satelite \cite{Klypin:1999uc,Moore:1999nt}, and the too-big-to-fail problem \cite{2011MNRAS.415L..40B}. These issues have motivated the exploration of alternative dark matter models, such as warm dark matter, self-interacting dark matter, fuzzy dark matter, condensate dark matter, and decaying dark matter. Most of these models involve modifications to the properties of dark matter, such as its mass, interactions, or decay channels, that lead to different predictions for the formation and evolution of structures in the Universe which can effectively be modeled within the Generalized Dark Matter framework. In this framework, various dark matter models can be effectively described by a dark matter component with a non-zero EoS parameter, which can be either constant or dynamical.

The measurements of BAO from second data release (DR2) of the Dark Energy Spectroscopic Instrument (DESI) based on three years of observations have been recently released \cite{DESI:2024mwx,DESI:2025fii,DESI:2025fxa,DESI:2025zgx,DESI:2025zpo}. The combination of DESI DR2 BAO data, CMB data, and SN data reveals a $2.8 \sigma$-$4.2 \sigma$ preference for dynamical dark energy within the Chevallier-Polarski-Linder (CPL) parameterization. This result has sparked significant interest in the investigation of alternative cosmological models that can potentially explain the DESI DR2 measurements. The joint analysis of DESI measurements with CMB and SNeIa favor the EoS of DE to be phantom-like ($w_{\rm DE}(z) < -1$) in the distant past and has since evolved to $w_{\rm DE}(z) > -1$ at present.

This preference has been observed in previous DESI analyses \cite{DESI:2024mwx,DESI:2024hhd,DESI:2024aqx,DESI:2024kob,DESI:2025fii} and continues even while allowing for variations in the spatial curvature \cite{DESI:2024mwx}, modified gravity \cite{Ishak:2024jhs}, or modifications to the pre-recombination physics \cite{Poulin:2024ken}. From a theoretical perspective, this phantom crossing becomes challenging to accommodate within the standard scalar field models of DE that are minimally coupled to gravity, as they are limited to satisfy $-1 \leq w_{\rm DE} \leq 1$. In particular, within the context of  general relativity (GR), a single field DE component with $w_{\rm DE} < -1$ would necessarily violate the null energy condition (NEC), given by $\rho + P \ge 0$ \cite{Hawking:1973uf}. While the phantom crossing can be seen as problematic due to stability issues, for instance,
within the framework of minimally coupled scalar fields,
obtaining such behavior is not difficult in extended theoretical frameworks. For example, phantom crossing can arise in models where DE possesses multiple
internal degrees of freedom, such as multi-field scenarios \cite{Hu:2004kh,Guo:2004fq,Wei:2005nw,Caldwell:2005ai,Cai:2007qw,Cai:2007zv}, non-standard vacuum models \cite{Parker:2000pr, Caldwell:2005xb},
frameworks with DE-DM interactions \cite{Amendola:1999er,Billyard:2000bh,Amendola:2003wa,Nojiri:2005sx,Clemson:2011an}, and modified theories of gravity \cite{Carvalho:2004ty,Hu:2007nk,Martin:2005bp,Anisimov:2005ne,Nesseris:2006jc,Shafieloo:2016bpk,Deffayet:2010qz,Pujolas:2011he,Ye:2024ywg,Wolf:2024stt}. Because of the aforementioned multiple internal degrees of freedom in these models, the effective, observable equation of state for the DE can cross the phantom divide even though the null energy condition is not violated. These considerations motivate exploring an alternative interpretation of the DESI preference, particularly that the observed deviation from $\Lambda$CDM may arise from unconventional DM dynamics rather than from a fundamentally dynamic DE component \cite{Chen:2025wwn, Giani:2025hhs,Li:2025dwz,Li:2025eqh}. The investigation of such a case in light of DESI DR2 measurements has revealed a mild preference for non-cold dark matter \cite{Kumar:2025etf}. Particularly, the DESI+DESY5 combination gives the DM equation of state to be $-0.084 \pm 0.035$, excluding CDM at 2.4$\sigma$ significance.

In this work, we explore the possibility of a dynamical dark matter (DDM) component with a non-zero EoS parameter as a viable interpretation of the recent measurements from the DESI DR2 data. We propose a phenomenological parameterization for the DM EoS which captures a smooth transition between two asymptotic regimes, allowing for a dynamical dark matter component that can deviate from the standard pressureless case. The article is organized as follows. In section \ref{sec:ddm}, we describe the theoretical framework for the DDM model and its implications for the background and perturbation equations. In section \ref{sec:data}, we specify the datasets and the methodology used for the analysis and parameter estimation. Section \ref{sec:results} deals with results and discussions of the constraints obtained on the DDM parameters and the preference of our model over $\Lambda$CDM and CPL parameterization. Finally, we conclude in section \ref{sec:conclusion} with a summary of our findings and future prospects for the DDM model.

\section{Dynamical Dark matter} \label{sec:ddm}

We consider a spatially flat Friedmann-Lemaitre-Robertson-Walker (FLRW) Universe filled with radiation, baryons, cosmological constant, and a dark matter component with energy density $\rho_{\rm dm}$ and EoS parameter $w_{\rm dm}$ \cite{Kumar:2012gr, Muller:2004yb,Armendariz-Picon:2013jej, Kopp:2018zxp}. In the presence of such dark matter component, the FLRW equations take the following form:
\begin{align}
    \mathcal{H}^{2}
    &= \frac{8 \pi G a^2}{3}
    \left( \rho_{\rm r} + \rho_{\rm b} + \rho_{\rm dm}
    + \rho_{\rm \Lambda}\right) \\
    \dot{\mathcal{H}}-\mathcal{H}^2
    &= -4 \pi G a^2
    \left( \frac{4}{3}\rho_{\rm r}  + (1+w_{\rm dm})\rho_{\rm dm} + \rho_b
    \right) 
\end{align}
Here $\rho_{\rm r,b}$ are the energy densities of radiation and baryonic fluid scaling as $a^{-4}$ and $a^{-3}$ respectively. The derivatives are with respect to conformal time and $\mathcal{H}\equiv \dot{a}/a$ is the Hubble parameter. The evolution of $\rho_{\rm dm}$, determined by solving the continuity equation, scales as
\begin{equation}
    \rho_{\rm dm} = \rho_{\text{dm},0}\, \exp \left( \int_{1}^{a} \, -\frac{3\,\left(1 + w_{\rm dm} (a')\right)}{a'} \,d a'\right)
\end{equation}
with $\rho_{\rm dm,0}$ being the current energy density. CMB observations and measurements of large-scale structure constrain the dark matter equation of state to remain close to pressureless behavior during recombination and early structure formation. Motivated by the preference of late-time cosmological data, including the DESI DR2 BAO measurements, for departures from $\Lambda$CDM, we investigate whether the dark matter equation of state can evolve to a non-zero value at low redshifts. For this, we adopt a logistic function of $\ln a$ as a minimal phenomenological parameterization that interpolates monotonically between these two regions. Hence, we parameterize the dark matter equation of state as
\begin{equation}\label{eq:w_form}
w_{\rm dm} = w_a^{\rm dm} + \frac{w^{\rm dm}_{0}-w^{\rm dm}_a}{1+(\frac{a_t}{a})^n}.
\end{equation}
This functional form describes a smooth transition between the early- and late-time asymptotic values of the dark matter EoS. For $n>0$, at early times ($a\ll a_t$), the term $(a_t/a)^n\gg1$, and the equation of state approaches $w_{\rm dm}\simeq w_a^{\rm dm}$. Conversely, at late times ($a\gg a_t$), $(a_t/a)^n\ll1$, such that $w_{\rm dm}\simeq w_0^{\rm dm}$. The present-day value of the EoS is obtained by setting $a=1$ in \cref{eq:w_form}. The parameter $a_t$ specifies the scale factor at which the EoS reaches the midpoint between its two asymptotic values, whereas the exponent $n$ controls the width of the transition in $\ln a$. This can be seen by writing the transition factor as $[1+\exp(-n\ln(a/a_t))]^{-1}$, showing that its characteristic width scales as $1/n$. Consequently, smaller values of $n$ yield a more gradual evolution over a wider range of scale factors, whereas larger values produce a sharper transition around $a_t$. This parameterization therefore allows the transition epoch and sharpness to be varied through $a_t$ and $n$, respectively.

Positive values of $w_a^{\rm dm}$ enhance the pre-recombination expansion rate by increasing early DM density, thereby reducing the comoving sound horizon at recombination ($r_\ast$) and shifting the acoustic angular scale ($\theta_\ast$) to smaller values. Since CMB observations tightly constrain $\theta_\ast$, this reduction in $r_\ast$ must be compensated by a decrease in the angular diameter distance $D_M$, which naturally favors a larger $H_0$. Thus, positive $w_a^{\rm dm}$ can provide a mechanism to partially alleviate the Hubble tension through early-time expansion modifications.

Following \cite{Ma:1995ey}, the perturbation
equations for the dark matter density contrast and velocity
divergence in the synchronous gauge can be respectively written as,
\begin{equation}
    \dot{\delta}_{\rm dm}
    = -(1+w_{\rm dm})
    \left(\theta_{\rm dm}+\frac{\dot{h}}{2}\right)
    -3\mathcal{H}
    \left(
    \frac{\delta p_{\rm dm}}{\delta\rho_{\rm dm}}
    -w_{\rm dm}
    \right)\delta_{\rm dm},
    \label{eq:delta_dm}
\end{equation}
\begin{equation}
    \dot{\theta}_{\rm dm}
    = -\mathcal{H}
    \left(1-3c_{\rm s,ad}^{2}\right)\theta_{\rm dm}
    +\frac{\delta p_{\rm dm}/\delta\rho_{\rm dm}}
    {1+w_{\rm dm}}k^{2}\delta_{\rm dm}
    -k^{2}\sigma_{\rm dm},
    \label{eq:theta_dm}
\end{equation}
The adiabatic sound speed is defined as
\begin{equation}
    c_{\rm s,ad}^{2}
    =\frac{\dot{p}_{\rm dm}}{\dot{\rho}_{\rm dm}}
    =w_{\rm dm}
    -\frac{\dot{w}_{\rm dm}}
    {3\mathcal{H}(1+w_{\rm dm})}.
    \label{eq:cs_ad}
\end{equation}
When the equation of state of a barotropic fluid becomes negative,
the adiabatic sound speed can become negative, potentially
leading to instabilities in the perturbations. To address this
issue, one can introduce an entropy perturbation and allow for
a non-negative effective sound speeds as well. Following \cite{Hu:1998kj}, the entropy perturbation is defined as,
\begin{equation}
    p_{\rm dm}\Gamma_{\rm dm}
    =\delta p_{\rm dm}
    -c_{\rm s,ad}^{2}\delta\rho_{\rm dm},
    \label{eq:entropy_perturbation}
\end{equation}
which is gauge invariant. In the dark matter rest frame, the
entropy perturbation is specified by
\begin{equation}
    w_{\rm dm}\Gamma_{\rm dm}
    =\left(c_{\rm s,eff}^{2}-c_{\rm s,ad}^{2}\right)
    \delta_{\rm dm}^{\rm rest},
    \label{eq:entropy_rest}
\end{equation}
where $c_{\rm s,eff}^{2}$ is the effective sound speed. Transforming
to an arbitrary gauge, in the dark matter rest frame, density contrast can be written as, \cite{Xu:2013mqe},
\begin{equation}
    \delta_{\rm dm}^{\rm rest}
    =\delta_{\rm dm}
    +3\mathcal{H}(1+w_{\rm dm})
    \frac{\theta_{\rm dm}}{k^{2}}.
    \label{eq:delta_rest}
\end{equation}
Using this relation the perturbation equations become,
\begin{multline}
    \dot{\delta}_{\rm dm} = -\left(1 + w_{\rm dm}\right) \, \left(\theta_{\rm dm} + \frac{\dot{h}}{2} \right) + \frac{\dot{w}_{\rm dm}}{1+w_{\rm dm}}\delta_{\rm dm} \\ - 3\,\mathcal{H} \left(c_{\rm s,eff}^{2} - c_{\rm s,ad}^{2}\right)
             \left[ \delta_{\rm dm} + 3\,\mathcal{H} \left(1 + w_{\rm dm}\right)\, \frac{\theta_{\rm dm}}{k^{2}}\right]
\end{multline}
\begin{multline}
    \dot{\theta}_{\rm dm} =  \frac{k^{2}  c_{\rm s,eff}^2}{1 + w_{\rm dm}}\, \left[ \delta_{\rm dm} + 3\,\mathcal{H} \left(1 + w_{\rm dm}\right)\, \frac{\theta_{\rm dm}}{k^{2}}\right] \\  -k^2\sigma_{\rm dm} -\mathcal{H}\theta_{\rm dm}
\end{multline}
Further, we assume the shear perturbation $\sigma_{\rm dm} = 0$. The effective sound speed $c_{\rm s,eff}^2$ characterizes DM's microscopic nature and determines the sound horizon ($l_s = c_{\rm s,eff}/H$), which separates scales where pressure suppresses clustering from those where gravitational clustering occurs. Smaller sound speeds allow DM perturbations to grow on smaller scales, enhancing their impact on structure formation and gravitational potential evolution, with further effects on the ISW effect in the CMB. Since efficient DM clustering is required to reproduce observed cosmic structure, we set $c_{\rm s,eff}^2=0$ \cite{Xu:2013mqe}.
%\textcolor{blue}{We do not claim a UV-complete realization here. The
%parametrization should be viewed as an effective description of a minimally coupled, separately conserved dark matter fluid, with dark energy given by a cosmological constant. A possible physical origin of the transition is bulk viscosity in the dark matter sector, which generates a negative effective pressure once the dark matter density falls below a critical value (see App.~\ref{app:viscous}); a microscopic derivation of this behavior is left for future work.}
\section{Data and Methodology}\label{sec:data}
We implement the theoretical model in a modified version of the Boltzmann solver code \texttt{CAMB} \cite{Lewis:1999bs,2012JCAP...04..027H}. We perform Markov-Chain-Monte Carlo (MCMC) simulations using the publicly available tool \texttt{COBAYA} \cite{Torrado:2020dgo,2019ascl.soft10019T}. Our model has 4 extra parameters $w_0^{\rm dm}$, $w_a^{\rm dm}$, $a_t$ and $n$, apart from the six standard cosmological parameters. We have added a superscript dm over the DDM parameters to discriminate them from the other standard parameters. We constrain them with uniform priors: the baryonic matter density $\Omega_b h^{2} \in [0.005, 0.1]$, the dark matter density $\Omega_{\rm dm}h^{2} \in [0.001,0.99]$, the logarithmic amplitude of primordial curvature spectrum $ {\rm{ln}}(10^{10} A_s) \in [1.6,3.9]$ evaluated at a suitable pivot scale, $k = 0.05 \  \text{Mpc}^{-1}$ along with its tilt $ n_s \in [0.8,1.2] $, the reionization optical depth $\tau_{\rm reio} \in [0.01,0.8] $, the angular scale of the acoustic fluctuations $ \theta_\ast \in [0.005,0.1] $ and our dynamical dark matter parameters $w_0^{\rm dm}, w_a^{\rm dm} \in [-0.1,0.1]$ and $a_t \in [0.0001,1]$. We have fixed the value of $n$ to be 3 in our baseline analysis. As discussed in Appendix~\ref{app:n_par}, when $n$ is varied over a broader prior range, its marginalized posterior becomes approximately flat for $n\gtrsim 2$, indicating that the data do not provide a robust constraint on this parameter. We therefore adopt $n=3$ as a representative choice in this weakly constrained region and examine the impact of varying it in the appendix. The priors on the DDM parameters $w_0^{\rm dm}, w_a^{\rm dm}$ are chosen so as to allow for moderate deviations from the standard CDM limit while remaining compatible with the current cosmological observations. In the standard scenario, DM is treated pressureless with $w_{\rm dm}=0$. Since these parameters correspond to the DM equation of state at different asymptotic limits, viable departures from this limit are expected to be small, since large positive or negative values of the DM equation of state would significantly alter the expansion history, matter clustering, and CMB anisotropies. In addition, allowing these parameters to take substantially more negative values would permit regimes in which the dark matter component exhibits dark energy like behavior, erasing the distinction between the dark matter and dark energy sectors. We use the standard three-neutrino description with a massive neutrino of mass $m_{\nu}$ = 0.06 eV, while the other two being  massless. The convergence of chains is ensured by having the Gelman-Rubin criterion $|R-1| \leq 0.05$ or the effective sample size becoming greater than $10^5$. We utilize  \texttt{GetDist} \cite{Lewis:2019xzd} and \texttt{BOBYQA} \cite{2018arXiv180400154C,2018arXiv181211343C} to analyze the chains and find the Maximum A Posteriori (MAP) $\chi^2_{\rm MAP}$, respectively. We calculate the differences in $\chi^2_{\rm MAP}$ and deviance information criterion (DIC) to get the preference of our model over $\Lambda$CDM. The DIC is calculated as 
\begin{equation}
    {\rm DIC} = 2\ \overline{\chi^{2}\left(\theta\right)} - \chi^{2}(\hat{\theta})
\end{equation}
with $\overline{\chi^{2}\left(\theta\right)}$ and $\chi^{2}(\hat{\theta})$ being the average of the effective $\chi^2$ over the posterior distribution and the best-fit $\chi^2$ respectively.
In our analysis, we use the following publicly available datasets  
\begin{enumerate}
        \item \textbf{CMB:-} We utilize the CMB likelihood built from four distinct components. First, the small-scale ($\ell > 30$) temperature and polarization power spectra, $C_{\ell}^{TT}$, $C_{\ell}^{TE}$, and $C_{\ell}^{EE}$, derived from the Planck \texttt{CamSpec} likelihood \cite{Planck:2018vyg, Efstathiou:2019mdh, Rosenberg:2022sdy}. Second, the large-scale ($2 \leq \ell \leq 30$) temperature spectrum, $C_{\ell}^{TT}$, obtained from the  Planck \texttt{Commander} likelihood \cite{Planck:2018vyg,Planck:2019nip}. Third, the large-scale ($2 \leq \ell \leq 30$) E-mode polarization spectrum, $C_{\ell}^{EE}$, is provided by the Planck \texttt{SimAll} likelihood~\cite{Planck:2018vyg,Planck:2019nip}. Finally, the CMB lensing likelihood from ACT DR6 is included \cite{ACT:2023kun}. We denote this likelihood by Pl.

        \item \textbf{DESI:-} We use the 13 DESI-BAO DR2 measurements across the redshift range $0.1 < z < 4.2$ obtained from observations of about 14 million galaxies and quasars which include bright galaxy sample (BGS), luminous red galaxies (LRG), emission line galaxies (ELG), quasars (QSO), and Lyman-$\alpha$ tracers \cite{DESI:2025zgx,DESI:2025zpo}. 
        
        \item \textbf{SNeIa:-} We use the PantheonPlus (PP) dataset \cite{Scolnic:2021amr} (1550 SNe, $0.001 < z < 2.26$), the Union3 compilation \cite{Rubin:2023ovl} (2087 SNe), and the recalibrated DES-Dovekie sample \cite{DES:2025sig,Popovic:2025glk}, a reanalysis of the original DES Y5 supernova survey incorporating improved photometric cross-calibration, updated white dwarf based zero points, and a retrained SALT3 light curve model. DES-Dovekie comprises 1623 likely DES Type Ia SNe and 197 SNe Ia at low redshift, drawn from the CfA3, CfA4, CSP, and Foundation samples, for a total of 1820 SNe.

        \item \textbf{Growth rate data:-} We use the robust and extended growth data set used to investigate the tension level between growth data measurements and $\Lambda$CDM model in the context of General Relativity \cite{Nesseris:2017vor}. The data set consists of $f\sigma_8$ measurements from  different surveys, which are displayed in appendix \ref{app:gold}.
         
\end{enumerate}
\begin{figure*}[t]
    \centering
    \includegraphics[width=0.47\textwidth]{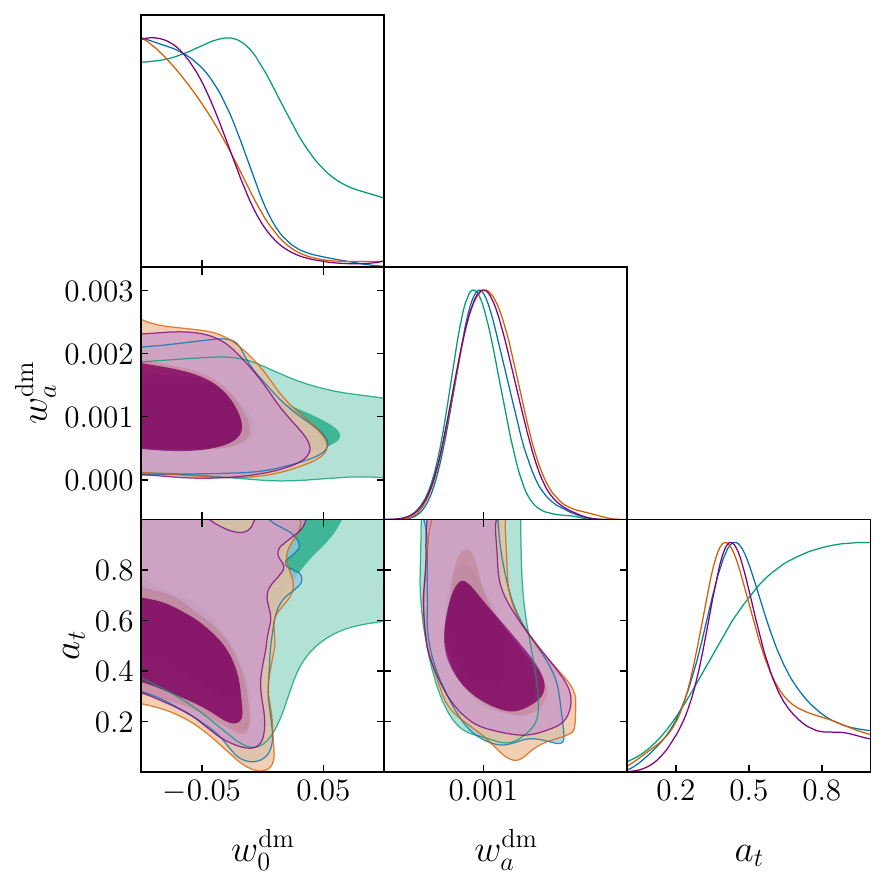}
    %\hfill
    \includegraphics[width=0.47\textwidth]{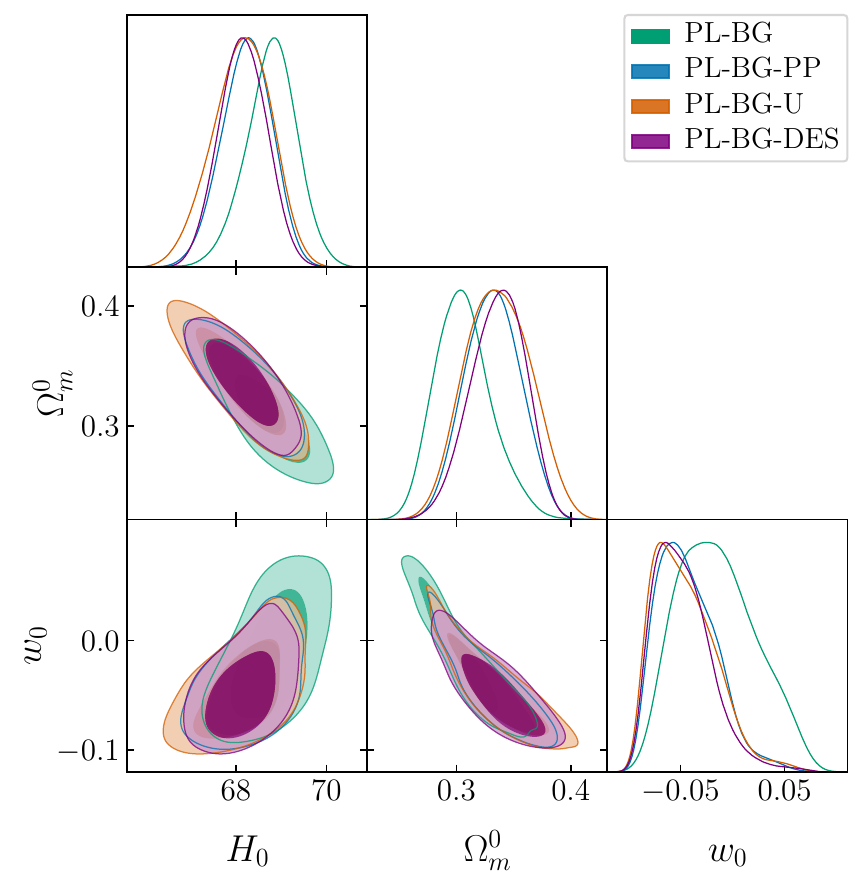}
    \caption{Corner plots of 1D and 2D marginalized posterior distributions for the DDM model, based on BAO from DESI DR2, CMB, supernovae and growth datasets. The left panel showcases the DDM parameters whereas the relevant derived parameters are displayed in the right panel. Contours at 68\% ($1\sigma$) and 95\% ($2\sigma$) levels showing parameter constraints and correlations within the framework.}
\label{fig:corner_dydn}
\end{figure*}
\section{Results and Discussion}\label{sec:results}
In this section, we present the constraints obtained on the cosmological parameters in our model. The marginalized parameter constraints along with their $1\sigma$ errors of the DDM model for different dataset combinations are reported in \cref{tab:dydn_const,tab:dydn_g17}. The constraints obtained from the $\Lambda$CDM and the CPL parameterization for similar dataset combinations are reported in appendix \ref{app:lcdm_cpl} for completeness. We begin our analysis by considering the combination of CMB and DESI DR2 data, which we denote by PL-B. We then add the PP, Union3, and DES-Dovekie SNeIa datasets to the PL-B combination, which we denote by PL-B-PP, PL-B-U, and PL-B-DES respectively. We also perform a separate analysis by including the growth data in the PL-B combination, which we denote by PL-BG using the similar notation for the combinations with the SNeIa datasets. We obtain the best-fit $\chi^2$ and DIC values for our model and compare them with the $\Lambda$CDM model to assess the preference of our model over $\Lambda$CDM as well as CPL parameterization. The differences in $\chi^2$ and DIC values are reported in \cref{tab:DIC}. 

\begin{figure*}[t]
    \centering    \includegraphics[width=0.95\linewidth,height=0.45\linewidth]{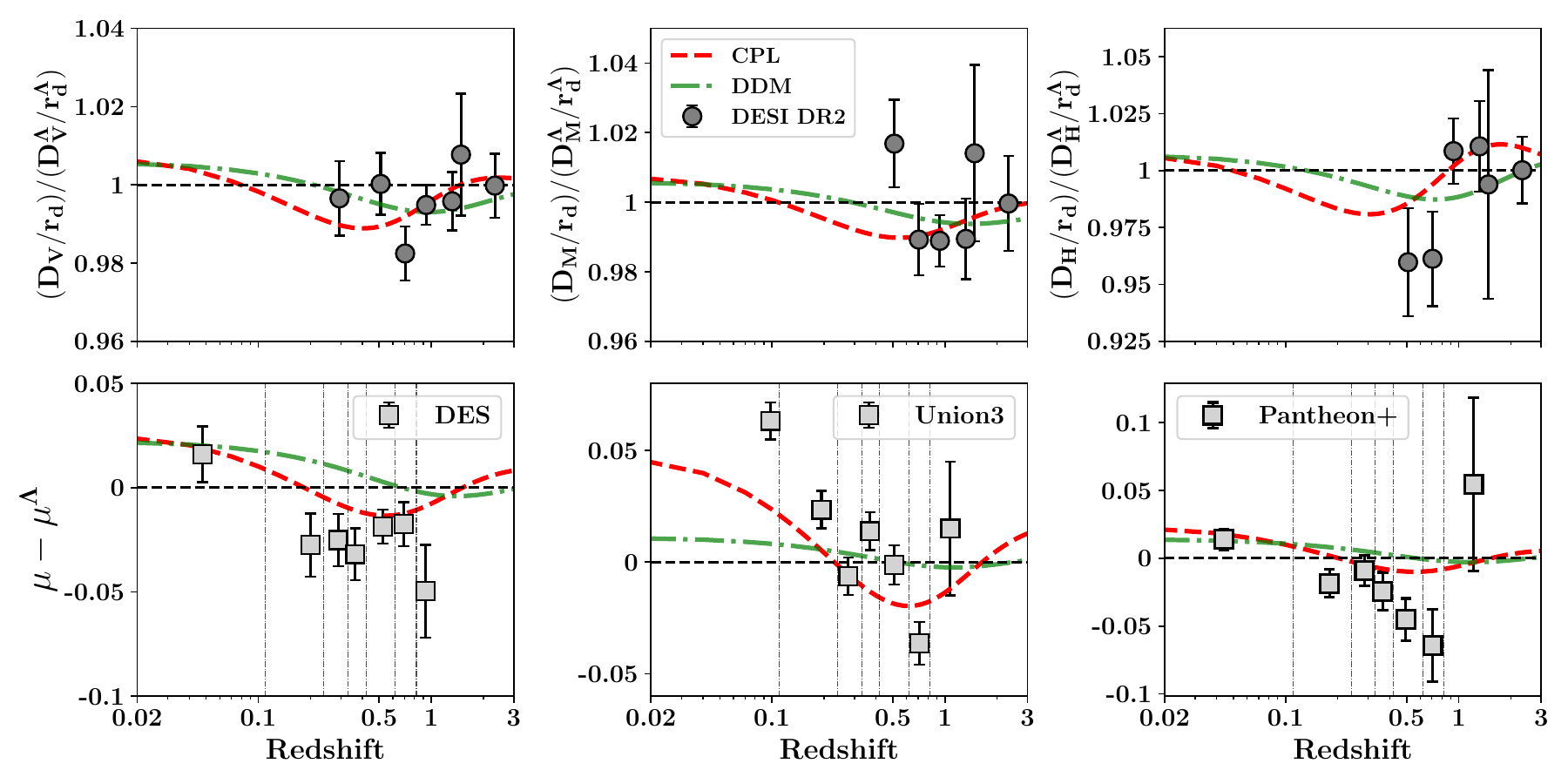}
    \caption{Hubble diagrams showing comparisons of DESI BAO and SNe data to models. In the top panels, DESI BAO measurements relative to the best fit $\Lambda$CDM parameters are shown with black circles, and DES is used as the SNe dataset for determining model fits. In the bottom panels, binned DES, Union3 and Pantheon+ SNe distance modulus residuals are shown with black squares. CPL and DDM residuals with respect to the corresponding best fit $\Lambda$CDM values are shown as red and green dotted lines respectively.}
    \label{fig:hubble}
\end{figure*}

\begin{table*}
\centering
%\scriptsize
\footnotesize
\begin{tabular}{|l|c|c|c|c|}
\hline
\hline
Parameter & \hspace{2em}PL-B\hspace{2em} & \hspace{2em}PL-B-DES\hspace{2em} & \hspace{2em}PL-B-PP\hspace{2em} & \hspace{2em}PL-B-U\hspace{2em} \\
\hline
$w_0^{\rm dm}$ & $-0.015^{+0.031}_{-0.076}$ & $< -0.0384$ & $< -0.0331$ & $< -0.0382$ \\

$w_a^{\rm dm}$ & $0.00097^{+0.00035}_{-0.00042}$ & $0.00113^{+0.00043}_{-0.00049}$ & $0.00109^{+0.00039}_{-0.00049}$ & $0.00116^{+0.00039}_{-0.00055}$ \\

$a_t$ & $> 0.510$ & $0.49^{+0.12}_{-0.23}$ & $0.52^{+0.16}_{-0.26}$ & $0.49^{+0.14}_{-0.25}$ \\

$H_0$ & $68.79^{+0.58}_{-0.52}$ & $68.14^{+0.53}_{-0.59}$ & $68.24\pm 0.57$ & $68.12^{+0.71}_{-0.63}$  \\

$\Omega_m$ & $0.305^{+0.021}_{-0.028}$ & $0.335^{+0.025}_{-0.023}$ & $0.330\pm 0.026$ & $0.336\pm 0.027$ \\

$\sigma_8$ & $0.810\pm 0.054$ & $0.754^{+0.035}_{-0.051}$ & $0.763^{+0.037}_{-0.060}$ & $0.754^{+0.040}_{-0.056}$ \\

$S_8$ & $0.814\pm 0.023$ & $0.794^{+0.017}_{-0.022}$ & $0.797^{+0.017}_{-0.025}$ & $0.795^{+0.018}_{-0.023}$ \\

$w_0$ & $-0.012^{+0.033}_{-0.046} \ (0.42)$ & $-0.045^{+0.020}_{-0.034} \ (1.66)$ & $-0.040^{+0.020}_{-0.040} \ (1.33)$ & $-0.044^{+0.020}_{-0.035} \ (1.52)$ \\

%$\chi^2$ &  &  &  &  \\
\bottomrule
\end{tabular}
\caption{The mean $\pm 1\sigma$ constraints on some of the relevant cosmological parameters inferred from various datasets including DESI DR2, CMB, and supernovae and their combinations for the dynamical dark matter model considered in this work. In the parentheses next to $w_0$, we report the equivalent Gaussian significance (in units of $\sigma$) of its deviation from zero, computed from the posterior tail probability derived from the MCMC chains. Here, $H_0$ is in units of km ${\rm s}^{-1}$ ${\rm Mpc}^{-1}$.}

\label{tab:dydn_const}

\end{table*}
\begin{table*}[t]
\centering
%\scriptsize
\footnotesize
\begin{tabular}{|l|c|c|c|c|}
\hline
\hline
Parameter & \hspace{1em}PL-BG\hspace{2em} & \hspace{2em}PL-BG-DES\hspace{2em} & \hspace{2em}PL-BG-PP\hspace{2em} & \hspace{2em}PL-BG-U\hspace{2em} \\
\hline
$w_0^{\rm dm}$ & $< 0.00115$ & $< -0.0436$ & $< -0.0359$ & $< -0.0401$ \\

$w_a^{\rm dm}$ & $0.00090^{+0.00037}_{-0.00043}$ & $0.00106^{+0.00043}_{-0.00052}$ & $0.00103^{+0.00038}_{-0.00051}$ & $0.00111^{+0.00044}_{-0.00055}$ \\

$a_t$ & $> 0.531$ & $0.50^{+0.11}_{-0.22}$ & $0.50^{+0.15}_{-0.23}$ & $0.48^{+0.13}_{-0.24}$ \\

$H_0$ & $68.79^{+0.59}_{-0.51}$ & $68.16\pm 0.53$ & $68.25^{+0.58}_{-0.52}$ & $68.12^{+0.71}_{-0.60}$  \\

$\Omega_m$ & $0.306^{+0.020}_{-0.027}$ & $0.337^{+0.026}_{-0.022}$ & $0.331\pm 0.024$ & $0.337\pm 0.028$ \\

$\sigma_8$ & $0.802\pm 0.052$ & $0.745^{+0.033}_{-0.050}$ & $0.754^{+0.038}_{-0.049}$ & $0.746^{+0.041}_{-0.058}$ \\

$S_8$ & $0.808\pm 0.022$ & $0.787^{+0.016}_{-0.021}$ & $0.791^{+0.017}_{-0.021}$ & $0.788^{+0.017}_{-0.024}$ \\

$w_0$ & $-0.015^{+0.030}_{-0.046} \ (0.45)$ & $-0.048^{+0.019}_{-0.033} \ (1.68)$ & $-0.043^{+0.022}_{-0.034} \ (1.51)$ & $-0.046^{+0.019}_{-0.037} \ (1.51)$ \\

\bottomrule
\end{tabular}
\caption{The mean $\pm 1\sigma$ constraints on  cosmological parameters inferred from the addition of growth data to the data combinations considered in \cref{tab:dydn_const}.
}
\label{tab:dydn_g17}
\end{table*}
From \cref{tab:dydn_const,tab:dydn_g17}, we find the constraints on the present day value of the DM EoS parameter $w_0$ deviating from zero at varying confidence levels for different dataset combinations. We see that $w_0$ becomes further negative when growth data is included. On the other hand, the DM EoS at early times is consistent with zero. Such a behaviour can arise naturally in effective viscous dark matter scenarios, where dissipative effects generate a small negative effective pressure at late times while preserving the standard pressure-less behaviour during the early Universe evolution. In this interpretation, the dark matter fluid behaves effectively as CDM at high redshifts, ensuring consistency with CMB and structure formation constraints, whereas viscous contributions become relevant only at late times and can lead to a mildly negative equation of state at present time. A theoretical realization of this correspondence to the viscous dark matter is discussed in appendix \ref{sec:VDM}. However, we do not claim a UV-complete realization here. A microscopic derivation of this behavior is left as a future work. Another key parameter is the transition scale factor $a_t$ which is also inferred at 68 \% CL. We present the 2 dimensional posterior distributions for the DDM parameters in \cref{fig:corner_dydn}. In \cref{fig:hubble} we show both the DESI DR2 BAO data for the distance measurements ($D_V$, $D_M$ and $D_H$) and the SNe data for the distance modulus $\mu$ against the best fits for the CPL and DDM models, all normalized to the best-fit $\Lambda$CDM constraints. For visualization, the SNe data have been binned in redshift and we have calculated the weighted average distance moduli and errors using the covariance matrix to include both statistical and systematic errors. We have used the CMB+BAO+SNeIa combination for the determination of residuals. We find that the transition scale factor $a_t$ is constrained to be greater than 0.4 at the 95 \% CL, indicating that the transition in the dark matter EoS occurs at relatively later times in the cosmic history. We obtain the present value of the Hubble parameter $H_0$ to be around 68 km ${\rm s}^{-1}$ ${\rm Mpc}^{-1}$ across all dataset combinations, which is consistent with $\Lambda$CDM predictions. The inferred value of $H_0$ is in tension with the local measurements from SH0ES \cite{Riess:2020fzl} at around 4$\sigma$ level. The matter density parameter $\Omega_m$ is found to be slightly higher than the $\Lambda$CDM predictions, with $\Omega_m = 0.335^{+0.025}_{-0.023}$ for the PL-B-DES combination. The error bars on $\Omega_m$ are also weaker as compared to $\Lambda$CDM and CPL models indicating a degeneracy between the DDM parameters and $\Omega_m$. The DDM model affects the growth of matter perturbations, which is characterized by the amplitude of matter fluctuations $\sigma_8$ and $S_8$ parameters. We find that $\sigma_8 = 0.754^{+0.035}_{-0.051}$ and $S_8 = 0.794^{+0.017}_{-0.022}$ for the PL-B-DES combination, which are almost $1\sigma$ lower than the $\Lambda$CDM predictions. The inclusion of growth data leads to a further decrease in the value of $\sigma_8$ and $S_8$, with $\sigma_8 = 0.745^{+0.033}_{-0.050}$ and $S_8 = 0.787^{+0.016}_{-0.021}$ for the PL-BG-DES combination, which is in better agreement with the weak lensing measurements from KiDS-1000 \cite{Hildebrandt:2016iqg} and DES-Y3 \cite{DES:2021wwk}.
\begin{table}[t]
%\centering
\hspace{-1 em}
\scriptsize
\scalebox{0.90}{
\begin{tabular}{|l|c|c|c|c|}
\hline
\hline
\textbf{Dataset} & $\bm{\Delta \chi^2_{\rm MAP}(\Lambda{\rm CDM})}$ & $\bm{\Delta \chi^2_{\rm MAP}({\rm CPL})} $ & $\bm{\Delta \mathrm{DIC}(\Lambda{\rm CDM})}$ & $\bm{\Delta \mathrm{DIC}({\rm CPL})}$ \\
\hline
\hline
PL-B             & $-6.558$ & $5.210$ & $-4.357$  & $3.199$ \\
PL-B-PP          & $-8.074$ & $1.773$ & $-5.233$  & $1.315$ \\
PL-B-U      & $-9.439$ & $7.676$ & $-4.971$  & $7.523$ \\
PL-B-DES       & $-10.975$& $2.975$ & $-3.905$  & $5.249$\\
PL-BG         & $-7.063$ & $4.272$ & $-1.930$  & $3.786$ \\
PL-BG-PP      & $-7.287$ & $2.166$ & $-3.520$  & $1.828$ \\
PL-BG-U  & $-9.224$ & $6.938$ & $-2.495$  & $9.118$ \\
PL-BG-DES   & $-7.944$& $4.169$ & $-3.427$  & $5.532$ \\
\hline
\hline
\end{tabular}
}
\caption{Comparison of dynamic dark matter model versus $\Lambda$CDM and CPL parameterization using different statistical tests: the maximum a posteriori $\chi^2$ difference ($\Delta \chi^2_{\rm MAP}$) and Deviance Information Criterion ($\Delta \mathrm{DIC}$). Negative $\Delta \chi^2_{\rm MAP}$ and $\Delta \mathrm{DIC}$ values favor the DDM model and positive values vice versa.
}
\label{tab:DIC}
\end{table}

Now, we discuss the preference of our model over $\Lambda$CDM and CPL parameterization. From \cref{tab:DIC}, we find that the DDM model provides a better fit to the data compared to $\Lambda$CDM, with a decrease in $\chi^2 \in [-6,-11]$ for different dataset combinations. The DIC values also indicate a preference for the DDM model over $\Lambda$CDM, with $\Delta {\rm DIC} \in [-2,-6]$ for different dataset combinations. The preference for the DDM model slightly decreases  when the growth data is included, which suggests that the $\Lambda$CDM model provides a better fit to the latter than the DDM model. When compared to the CPL model, we find the CPL model is preferred over the DDM model as shown by the positive $\Delta \chi^2$ and $\Delta {\rm DIC}$ values. This is expected since the CPL model has more freedom in the dark energy sector, which can lead to a better fit to the data.
\begin{figure}
    \centering
    \includegraphics[width=\linewidth,height=0.70
    \linewidth]{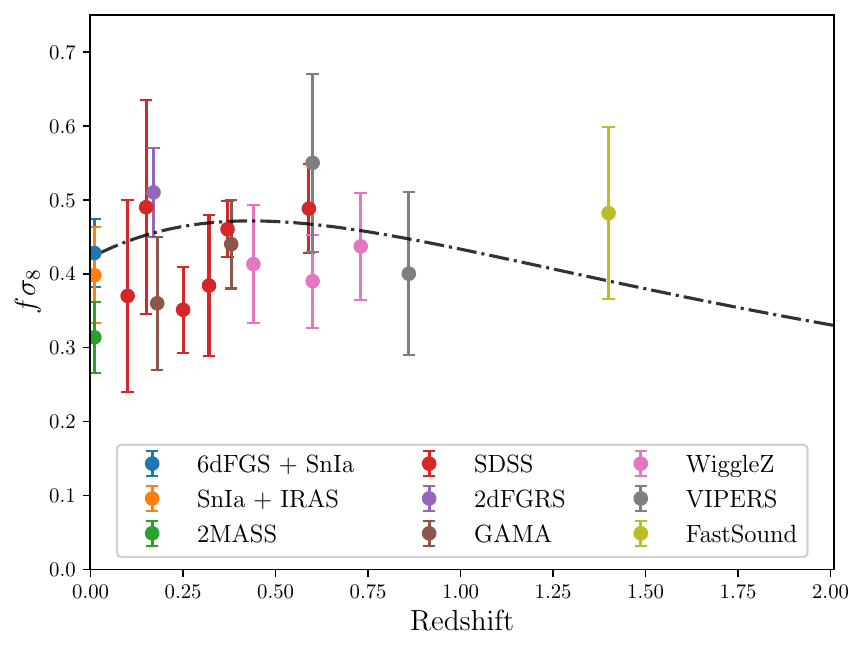}
    \caption{Reconstruction of $f\sigma_8(z)$ obtained from CMB+DESI+DES+G17 constraints (shown by the dashed line) for DDM model. The $f\sigma_8$ measurements used for obtaining them are plotted for reference.}
    \label{fig:growth}
\end{figure}

\section{Conclusion}\label{sec:conclusion} 
The recent DESI DR2 BAO measurements, in combination with CMB and SNeIa data, have provided a persistent preference for a NEC violating, dynamical dark energy component described by the CPL parameterization. Since such phantom crossing behaviour is difficult to accommodate within minimally coupled scalar field models without invoking extended theoretical frameworks, in this work we explored an alternative interpretation in which the departure from $\Lambda$CDM originates from unconventional dark matter dynamics rather than a genuinely dynamical DE sector. To this end, we introduced a dynamical dark matter model in which the equation of state is no longer fixed at $w_{\rm dm}=0$, but instead allowed to evolve smoothly between two asymptotic values, $w_a^{\rm dm}$ at early times and $w_0^{\rm dm}$ at late times, through the phenomenological parameterization introduced in \cref{eq:w_form}. This construction is characterized by four additional parameters, $w_0^{\rm dm}$, $w_a^{\rm dm}$, $a_t$, and $n$, on top of the six standard $\Lambda$CDM parameters, and modifies both the background expansion history and the linear perturbation equations of the DM fluid. For this, we implemented the DDM model in a modified version of \texttt{CAMB} and performed MCMC parameter estimation with \texttt{COBAYA}, using the CMB likelihoods built from Planck \texttt{CamSpec}, \texttt{Commander}, and \texttt{SimAll}, together with ACT DR6 CMB lensing, DESI DR2 BAO measurements, and three independent SNeIa compilations (PantheonPlus, Union3, and DES-Dovekie). We further examined the impact of including an independent compilation of $f\sigma_8$ growth rate measurements, appropriately corrected for the fiducial cosmology dependence of the reported values, in order to test the robustness of our results against LSS growth data.

Our main result is that the dark matter EoS at early times, $w_a^{\rm dm}$, remains fully consistent with the standard pressure-less CDM limit across all dataset combinations, safeguarding the successful predictions of $\Lambda$CDM for CMB anisotropies and structure formation in the early Universe. In contrast, the present day value of the EoS is found to be negative and consistent with 0 only marginally, with a statistical significance ranging from about $0.42\sigma$ to $1.66\sigma$ depending on the CMB, BAO and SNeIa combination considered. The strongest preference for this is obtained for the PL-B-DES combination, which yields the present day EoS to be $-0.045^{+0.020}_{-0.034}$, together with a transition scale factor $a_t = 0.49^{+0.12}_{-0.23}$ at 68\% CL, indicating that the departure from CDM occurs at relatively low redshifts. This preference for a negative EoS persists, and in fact becomes slightly stronger, once growth rate data are incorporated into the analysis, indicating that our results are not an artefact of any single dataset but rather a consistent feature shared by the geometric (BAO, SNeIa) and dynamical (growth) probes considered here.

The inclusion of a non-zero, time-varying dark matter EoS also leaves a distinct imprint on the derived cosmological parameters. We obtain $H_0 \simeq 68\ {\rm km\,s^{-1}\,Mpc^{-1}}$ across all dataset combinations, consistent with the $\Lambda$CDM value and still in tension with the SH0ES local measurement at about $4\sigma$ level, indicating that the DDM model in its current form does not resolve the Hubble tension. The matter density $\Omega_m$ is shifted to slightly higher values relative to $\Lambda$CDM, with correspondingly weaker error bars, signalling a degeneracy between $\Omega_m$ and the DDM parameters. More notably, the amplitude of matter fluctuations is suppressed relative to $\Lambda$CDM, with $\sigma_8 = 0.754^{+0.035}_{-0.051}$ and $S_8 = 0.794^{+0.017}_{-0.022}$ for the PL-B-DES combination, slightly decreasing further to $\sigma_8 = 0.745^{+0.033}_{-0.050}$ and $S_8 = 0.787^{+0.016}_{-0.021}$ once growth data are included. These values lie closer to the weak-lensing determinations of $S_8$ from KiDS-1000 and DES-Y3, suggesting that the DDM model, in addition to reinterpreting the DESI preference, can simultaneously provide some relief to the longstanding $S_8$ tension.

We quantified the statistical preference for the DDM model relative to $\Lambda$CDM and to the CPL parameterization of dynamical dark energy using both $\Delta \chi^2_{\rm MAP}$ and $\Delta {\rm DIC}$ (\cref{tab:DIC}). The DDM model consistently provides a better fit to the data than $\Lambda$CDM, with $\Delta \chi^2_{\rm MAP}$ ranging between $-6.6$ and $-10.975$ and $\Delta {\rm DIC}$ between $-1.9$ and $-5.233$ across the dataset combinations considered, indicating a moderate to strong preference for the extra DM freedom even after accounting for the additional parameters through the DIC. This preference is somewhat reduced, though not eliminated, once growth rate data are added, suggesting a mild tension between the LSS growth measurements and the specific late time suppression of structure favoured by the DDM model. When compared against the CPL parameterization, however, the DDM model is disfavoured, with positive $\Delta \chi^2_{\rm MAP}$ and $\Delta {\rm DIC}$ values throughout. This is expected given that the CPL model has an extra degree of freedom in the DE sector and is specifically tailored to capture the phantom crossing behaviour favoured by the DESI data. Nevertheless, the fact that a comparatively simple, four-parameter modification of the DM sector alone can reproduce a sizeable fraction of the improvement in fit obtained with a dynamical DE component is a non-trivial finding, and reinforces the interpretation that the DESI anomaly need not necessarily be attributed to new physics in the DE sector.

Physically, the qualitative behaviour we find, a DM EoS that is negligible at early times but becomes mildly negative at late times, is naturally realized in effective viscous DM scenarios, where bulk viscous dissipation generates a small negative effective pressure that only becomes relevant once the Universe has sufficiently diluted the DM density, while leaving the early time, pressure-less behaviour of CDM intact. We demonstrate this correspondence explicitly in appendix \ref{sec:VDM}, where a simple bulk viscosity model with a density dependent viscosity coefficient reproduces the same asymptotic transition in $w_{\rm dm}$ found in our phenomenological parameterization, thereby providing physical motivation for the functional form adopted in \cref{eq:w_form} and connecting our purely observational analysis to an underlying microphysical origin.

Taken together, our results show that a dynamical dark matter component provides a viable and statistically competitive alternative to dynamical dark energy in interpreting the DESI DR2 preference for deviations from $\Lambda$CDM, while also offering a possible avenue toward alleviating the $S_8$ tension, without requiring any exotic scalar field or NEC violating physics in the dark energy sector. Future work could extend this analysis by considering a dynamical dark energy component alongside DDM and allowing for interactions between the two sectors. Such interactions could modify both the background expansion and the evolution of cosmological perturbations, potentially introducing degeneracies with the effects specific to a time-dependent dark matter EoS. Furthermore, in specific microscopic models involving fermionic dark matter coupled to a dynamical dark energy field, interactions and possible condensation effects may modify the effective equation of state and allow the effective dark energy to cross the phantom divide without necessarily requiring the underlying dark energy to behave as a phantom field. Further extensions could also explore degeneracies between the DDM parameters and the neutrino sector or the effective sound speed $c_s^2$, and by testing the model against forthcoming DESI data releases, weak-lensing surveys such as Euclid and LSST, and independent redshift-space distortion measurements, which will allow for a more stringent test of the late time suppression of structure growth predicted by this class of models.

\appendix
\section{Gold growth data}\label{app:gold} 
In this appendix, we provide details of the compilation of $f\sigma_8$ growth data used in this work. The data are taken from \cite{Nesseris:2017vor} and are summarized in \cref{tab:gold_data}. 
\begin{table*}[b]
\centering
\label{tab:gold_data}
\small
\setlength{\tabcolsep}{4pt}

\begin{tabular}{|l|ccc@{\qquad}|l|ccc|}
\hline\hline
Dataset & $z$ & $f\sigma_8$ & $\Omega_m^{\rm fid}$ &
Dataset & $z$ & $f\sigma_8$ & $\Omega_m^{\rm fid}$ \\
\hline
6dFGS + SnIa & 0.02 & $0.428\pm0.047$ & 0.30 &
GAMA & 0.38 & $0.440\pm0.060$ & 0.27 \\
SnIa + IRAS & 0.02 & $0.398\pm0.065$ & 0.30 &
WiggleZ & 0.44 & $0.413\pm0.080$ & 0.27 \\
2MASS & 0.02 & $0.314\pm0.048$ & 0.266 &
SDSS-CMASS & 0.59 & $0.488\pm0.060$ & 0.307 \\
SDSS-veloc & 0.10 & $0.370\pm0.130$ & 0.30 &
WiggleZ & 0.60 & $0.390\pm0.063$ & 0.27 \\
SDSS-MGS & 0.15 & $0.490\pm0.145$ & 0.31 &
Vipers PDR-2 & 0.60 & $0.550\pm0.120$ & 0.30 \\
2dFGRS & 0.17 & $0.510\pm0.060$ & 0.30 &
WiggleZ & 0.73 & $0.437\pm0.072$ & 0.27 \\
GAMA & 0.18 & $0.360\pm0.090$ & 0.27 &
Vipers PDR-2 & 0.86 & $0.400\pm0.110$ & 0.30 \\
SDSS-LRG-200 & 0.25 & $0.351\pm0.058$ & 0.25 &
FastSound & 1.40 & $0.482\pm0.116$ & 0.270 \\
BOSS-LOWZ & 0.32 & $0.384\pm0.095$ & 0.274 &
 & & & \\
\hline\hline
\end{tabular}
\caption{Compilation of robust and independent $f\sigma_8$ measurements.
The corresponding fiducial matter density is given by $\Omega_m^{\rm fid}$.}
\end{table*}
Since the data have a dependence on the fiducial model used by the collaborations to convert redshifts to distances, a correction has to be implemented. For this, we define the ratio the product of the Hubble parameter $H(z)$ and the angular diameter distance $d_A(z)$ for the model at hand to that of the fiducial cosmology.
        \begin{equation}
            r(z) = \frac{H(z)d_A(z)}{H^{\rm fid}(z)d_A^{\rm fid}(z)}
        \end{equation}
        We can define a vector $V^i(z_i,p^j)$,
        \begin{equation}
            V^i(z_i,p^j) = f\sigma_{8,i}-r(z_i)f\sigma_{8,i}(z_i,p^j)
        \end{equation}
        where $p^j$ is the $j$th component of the vector containing the cosmological parameters. $f\sigma_{8,i}$ is the value of the $i$th data point,  which we want to determine, while $f\sigma_{8,i}(z_i,p^j)$ is the theoretical prediction, both at redshift $z_i$. Hence, 
        \begin{equation}
            \chi^2_{\rm growth} = V^iC_{ij}^{-1}V^j
        \end{equation}
        where $C_{ij}$ is the covariance matrix of the data. Following the analysis in \cite{Nesseris:2017vor}, we assume most of the data are not correlated, except the ones from WiggleZ, where the covariance matrix is given as \cite{2012MNRAS.425..405B},
        \begin{equation}
            C_{ij}^{\rm WiggleZ}=10^{-3}\begin{pmatrix}
                6.400 & 2.570 & 0.000 \\
                2.570 & 3.969 & 2.540 \\
                0.000 & 2.540 & 5.184
            \end{pmatrix}
        \end{equation}
\begin{figure}
    \centering
    \includegraphics[width=\linewidth]{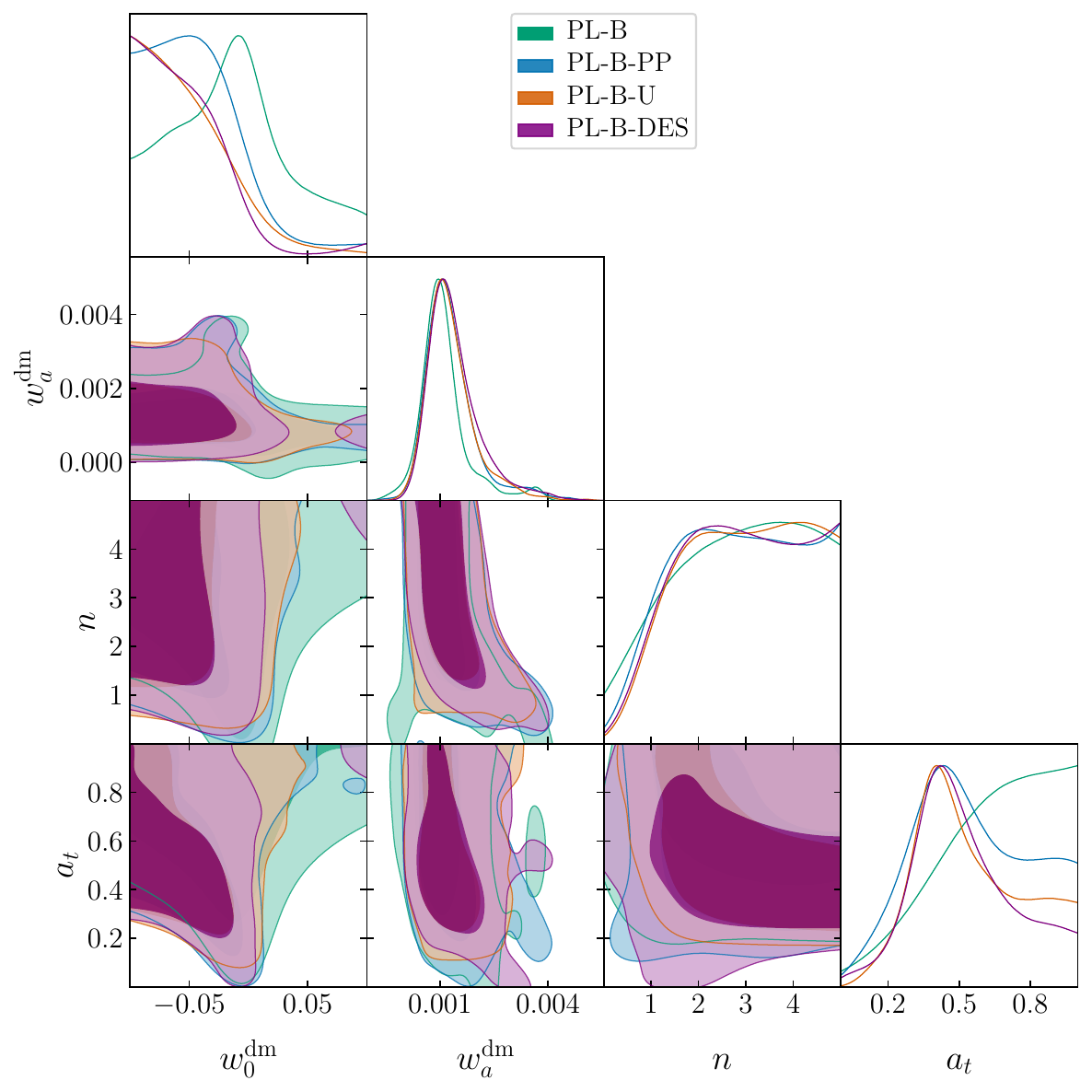}
    \caption{Corner plots of 1D and 2D marginalized posterior distributions for the DDM model, based on BAO from DESI DR2, CMB, supernovae and growth datasets. Contours at 68\% ($1\sigma$) and 95\% ($2\sigma$) levels showing parameter constraints and correlations within the framework.}
    \label{fig:corner_ndm}
\end{figure}
\section{Effect of the parameter $n$} \label{app:n_par}
In this appendix, we investigate the more general case in which the exponent $n$ in the dark matter EoS parameterization is treated as a free parameter, rather than being fixed to $n=3$ as in our baseline analysis. This parameter controls the sharpness of the transition between the early-time and late-time asymptotic values of the dark matter EoS. We constrain the model using DESI DR2 BAO measurements in combination with CMB and supernovae data, adopting a flat prior $n\in[0.0001,5]$. The one and two dimensional marginalized posterior distributions are shown in \cref{fig:corner_ndm}.

We find that the posterior distribution rises from small values of $n$ and becomes approximately flat for $n\gtrsim 2$. This behavior indicates that the data do not provide a robust constraint on $n$, particularly in the region $n\gtrsim 2$. Consequently, it will not be able to identify a preferred value of $n$ within this region. This weak constraint motivates fixing $n$ to a representative value in the baseline analysis, thereby reducing the dimensionality of the parameter space without imposing a data-driven preference for the parameter $n$.

To assess the impact of allowing $n$ to vary, we compare the differences in the maximum-a-posteriori (MAP) chi-squared statistic, $\Delta\chi^2_{\rm MAP}$, relative to the $\Lambda$CDM and CPL parameterizations for the same combinations of cosmological datasets. These results are presented in \cref{tab:del_xi_ndm}. We find that the $\Delta\chi^2_{\rm MAP}$ values obtained when $n$ is varied are not substantially different from those obtained when $n$ is fixed to a representative value within the approximately flat region of its posterior. This suggests that allowing $n$ to vary does not significantly improve the fitting of the dynamical dark matter model relative to the reference parameterizations. Hence, the main conclusions of our baseline analysis are not sensitive to whether $n$ is fixed or varied within the range explored.

\begin{table}[h!]
%\centering
\hspace{-1 em}
\scriptsize
\scalebox{1.2}{
\begin{tabular}{|l|c|c|}
\hline
\hline
\textbf{Dataset} & $\bm{\Delta \chi^2_{\rm MAP}(\Lambda{\rm CDM})}$ & $\bm{\Delta \chi^2_{\rm MAP}({\rm CPL})} $  \\
\hline
\hline
PL-B          & $-7.272$ & $4.496$  \\
PL-B-PP       & $-8.801$ & $1.046$  \\
PL-B-U        & $-9.560$ & $7.555$  \\
PL-B-DES      & $-9.859$& $4.091$  \\
\hline
\hline
\end{tabular}
}
\caption{Comparison of dynamic dark matter model versus $\Lambda$CDM and CPL parameterization using the maximum a posteriori $\chi^2$ difference ($\Delta \chi^2_{\rm MAP}$).
}
\label{tab:del_xi_ndm}
\end{table}

\begin{figure*}[t]
    \centering    \includegraphics[width=0.9\linewidth,height=0.3\linewidth]{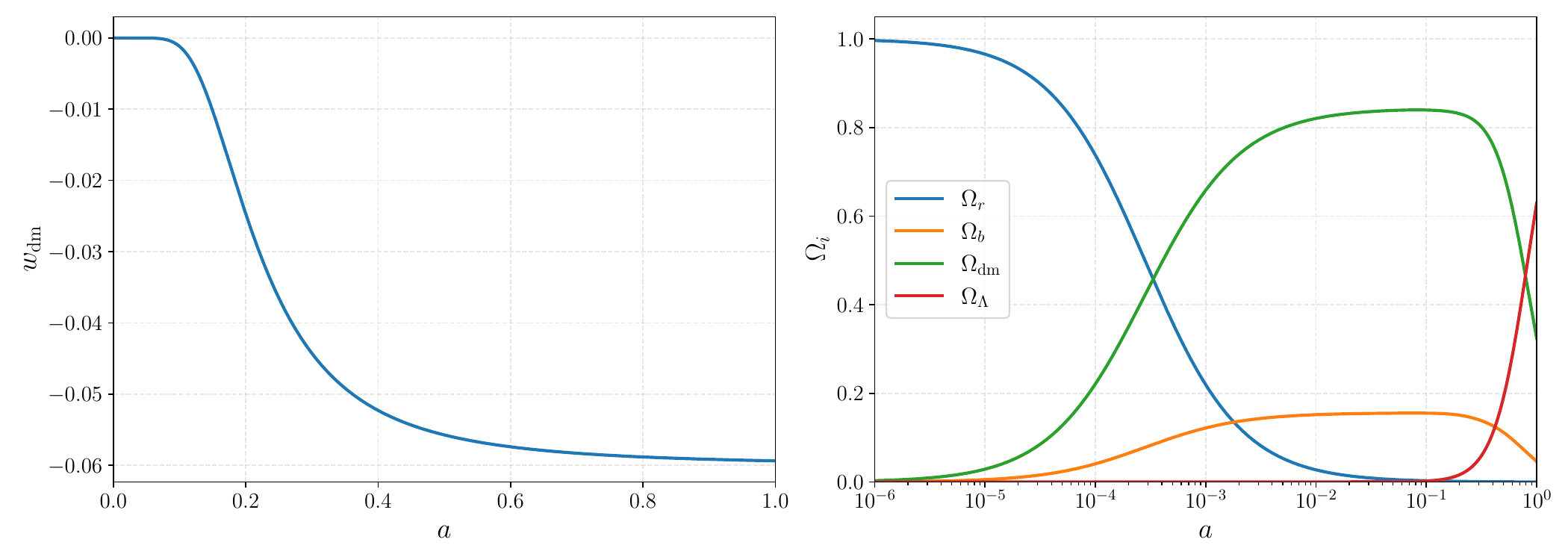}
    \caption{The left panel shows the evolution of the EoS parameter while the right panel displays the evolution for the density parameters against the scale factor $a$. The model parameters $\rho_c$, $m$ and $b$ are taken to be 100, 3 and 0.02 respectively.}
    \label{fig:O_w_evo}
\end{figure*}
\begin{figure*}[b]
    \centering    \includegraphics[width=0.9\linewidth,height=0.3\linewidth]{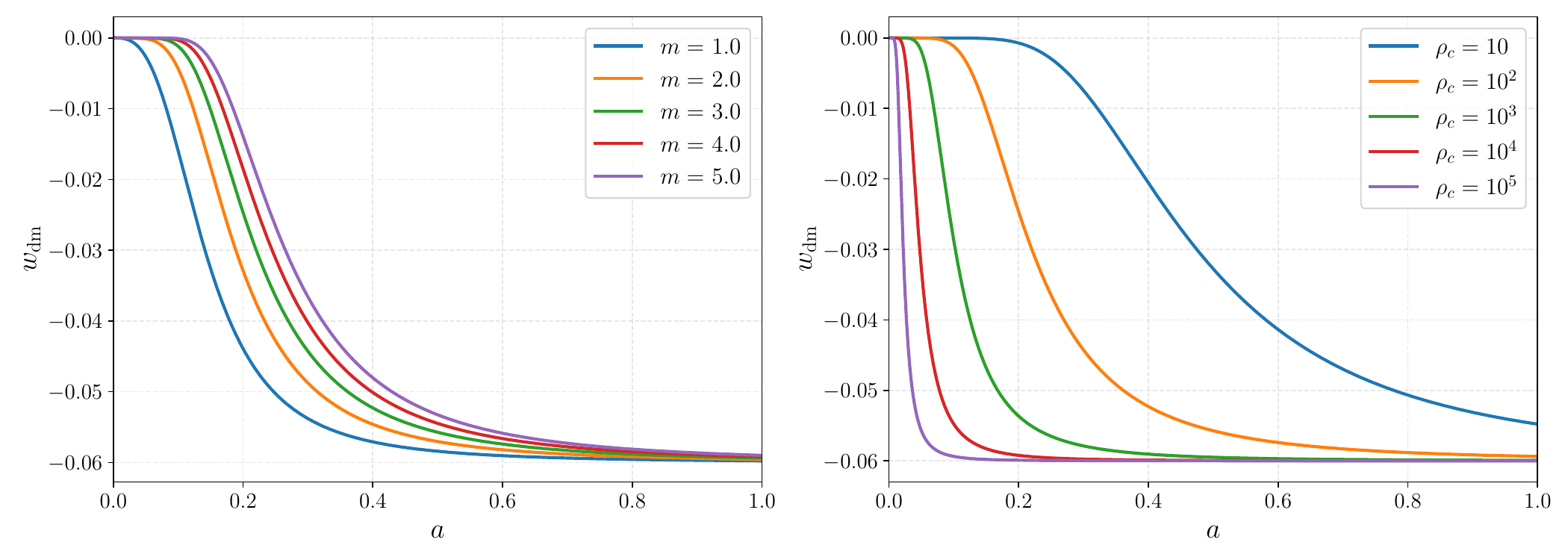}
    \caption{Dependence of $w_{\rm dm}$ on the model parameters $m$ and $\rho_c$. In the left panel, $\rho_c$ is taken to be 100, while the value of $m$ is kept at 3 in the right panel. $b$ is taken to be 0.02.}
    \label{fig:w_dep}
\end{figure*}
\section{Viscous dark matter}\label{sec:VDM}
\noindent We model dark matter as a fluid equipped with non-zero bulk viscosity. In the presence of bulk viscosity, dissipative effects modify the effective pressure as
\begin{equation}
    p_{\rm eff} = p - 3H\zeta \, ,
\end{equation}
where $p$ denotes the equilibrium pressure, $H$ is the Hubble expansion rate, and $\zeta$ is the bulk viscosity coefficient. For cold dark matter ($p=0$), the effective pressure reduces to
\begin{equation}
    p_{\rm eff} = - 3H\zeta \, ,
\end{equation}
which corresponds to a negative contribution arising from viscous dissipation. With this the effective equation of state of DM becomes \cite{Anand:2017wsj},
\begin{equation}
    w_{\rm dm} = -\frac{3H\zeta}{\rho_{\rm dm}}
\end{equation}
We choose the viscosity coefficient $\zeta$ to take the following functional form,
\begin{equation}
%\boxed{
\zeta(\rho_{\rm dm}, H)
= \left( \frac{\rho_c}{\rho_{\rm dm} + \rho_c} \right)^m
\frac{\rho_{\rm dm} \ b}{H}
%}.
\end{equation}
where $\rho_c$, $m$ and $b$ are constant parameters controlling the evolution of the DM fluid. With this functional form, the expression for the DM equation of state now becomes,
\begin{equation}
%\boxed{
w_{\rm dm}(\rho_{\rm dm})
= -3b \left(
\frac{\rho_c}{\rho_{\rm dm} + \rho_c}
\right)^m
%}.
\end{equation}
In the limit, $\rho_{\rm dm} \gg \rho_c$, $w_{\rm dm}\sim 0$, while the opposite limit yields $w_{\rm dm}\sim -3b$.  
The dark matter energy conservation equation is
\begin{equation}
\dot{\rho}_{\rm dm}
+ 3H \left(\rho_{\rm dm} + p_{\rm bulk} \right) = 0,
\end{equation}
with $p_{\rm bulk} = -3H\zeta$. Switching to $N = \ln a$, we obtain,
\begin{eqnarray}
    \frac{d\rho_{\rm dm}}{dN}
&=& -3\left( 1 + w_{\rm dm}(\rho_{\rm dm}) \right)
\rho_{\rm dm} \\ &=& -3\left[
1 - 3b
\left( \frac{\rho_c}{\rho_{\rm dm} + \rho_c} \right)^m
\right]
\rho_{\rm dm}
%}.
\end{eqnarray}
Solving the continuity equation along with the evolution of other components, we obtain the evolution of the density parameters ($\Omega_i$) which is depicted in \cref{fig:O_w_evo} along with the evolution of the EoS parameter. The parameters $\rho_c$, $m$, and $b$ are phenomenological parameters that characterize the density dependence and the magnitude of the viscous contribution. The parameter $\rho_c$ determines the characteristic density at which the transition from nearly pressureless behavior to a viscous regime occurs, while $m$ controls the sharpness of this transition. The dimensionless parameter $b$ sets the asymptotic value of the EoS parameter in the low density limit. In particular, increasing $b$ leads to a more negative asymptotic equation of state corresponding to late times. The dependence of $w_{\rm dm}$ on the model parameters $\rho_c$ and $m$ is displayed in \cref{fig:w_dep}. The values adopted in our figures are chosen to illustrate the qualitative behavior of the model and are not derived from model-independent theoretical or observational constraints.

%\newpage

\section{Cosmological constraints}\label{app:lcdm_cpl}
In this appendix, we display the full parameter constraints obtained from the dataset combinations including CMB, BAO, Supernovae and growth data for the dynamical dark matter model, $\Lambda$CDM model and CPL parameterisation.

\begin{table*}[h]
\centering
\scriptsize
%\footnotesize
\renewcommand{\arraystretch}{0.5}
\begin{tabular}{|l c c c c|}
\hline
\hline
Parameter & \hspace{2em}PL-B\hspace{2em} & \hspace{2em}PL-B-DES\hspace{2em} & \hspace{2em}PL-B-PP\hspace{2em} & \hspace{2em}PL-B-U\hspace{2em} \\
\hline
\multicolumn{5}{|l|}{\textit{\textbf{Sampled}}} \\

$100\theta_{\ast}$ & \hspace{2em}$1.04070^{+0.00063}_{-0.00055}$\hspace{2em} & \hspace{1em}$1.04009^{+0.00049}_{-0.00060}$\hspace{1em} & \hspace{1em}$1.04019^{+0.00056}_{-0.00064}$\hspace{1em} & \hspace{1em}$1.04011^{+0.00056}_{-0.00064}$\hspace{1em}  \\

$\Omega_b h^2$ & $0.02215^{+0.00013}_{-0.00015}$ & $0.02215\pm 0.00014$ & $0.02215\pm 0.00014$ & $0.02214\pm 0.00014$  \\

$\Omega_c h^2$ & $0.1214^{+0.0087}_{-0.011}$ & $0.133^{+0.010}_{-0.0085}$ & $0.131^{+0.011}_{-0.0094}$ & $0.133\pm 0.010$ \\

$\log(10^{10}A_s)$ & $3.046\pm 0.014$ & $3.047\pm 0.014$ & $3.047\pm 0.014$ & $3.047\pm 0.013$ \\

$n_s$ & $0.9658\pm 0.0037$ & $0.9660\pm 0.0038$ & $0.9660\pm 0.0037$ & $0.9662\pm 0.0038$ \\

$\tau_{\rm reio}$ & $0.0549\pm 0.0073$ & $0.0553\pm 0.0077$ & $0.0552\pm 0.0073$ & $0.0552\pm 0.0071$ \\

$w_0^{\rm dm}$ & $-0.015^{+0.031}_{-0.076}$ & $< -0.0384$ & $< -0.0331$ & $< -0.0382$ \\

$w_a^{\rm dm}$ & $0.00097^{+0.00035}_{-0.00042}$ & $0.00113^{+0.00043}_{-0.00049}$ & $0.00109^{+0.00039}_{-0.00049}$ & $0.00116^{+0.00039}_{-0.00055}$ \\

$a_t$ & $> 0.510$ & $0.49^{+0.12}_{-0.23}$ & $0.52^{+0.16}_{-0.26}$ & $0.49^{+0.14}_{-0.25}$ \\
\hline
\multicolumn{5}{|l|}{\textit{\textbf{Derived}}} \\

$H_0$ & $68.79^{+0.58}_{-0.52}$ & $68.14\pm 0.53$ & $68.24\pm 0.57$ & $68.12^{+0.71}_{-0.63}$  \\

$\Omega_m$ & $0.305^{+0.021}_{-0.028}$ & $0.335^{+0.025}_{-0.023}$ & $0.330\pm 0.026$ & $0.336\pm 0.027$ \\

$\sigma_8$ & $0.810\pm 0.054$ & $0.754^{+0.035}_{-0.051}$ & $0.763^{+0.037}_{-0.060}$ & $0.754^{+0.040}_{-0.056}$ \\

$S_8$ & $0.814\pm 0.023$ & $0.794^{+0.017}_{-0.022}$ & $0.797^{+0.017}_{-0.025}$ & $0.795^{+0.018}_{-0.023}$ \\

$w_0$ & $-0.012^{+0.033}_{-0.046} $ & $-0.045^{+0.020}_{-0.034} $ & $-0.040^{+0.020}_{-0.040} $ & $-0.044^{+0.020}_{-0.035} $ \\

%$\chi^2$ &  &  &  &  \\
\bottomrule
\end{tabular}
\caption{The mean $\pm 1\sigma$ constraints on  cosmological parameters inferred from various datasets including DESI DR2, CMB, supernovae and their combinations for the dynamical dark matter model considered in this work.}
%\label{tab:dydn_const}

\end{table*}

\begin{table*}[t]
\centering
\scriptsize
%\footnotesize
\renewcommand{\arraystretch}{0.5}
\begin{tabular}{|lcccc|}
\hline
\hline
Parameter & \hspace{2em}PL-BG\hspace{2em} & \hspace{2em}PL-BG-DES\hspace{2em} & \hspace{2em}PL-BG-PP\hspace{2em} & \hspace{2em}PL-BG-U\hspace{2em} \\
\hline
\multicolumn{5}{|l|}{\textit{\textbf{Sampled}}} \\

$100\theta_{\ast}$ & \hspace{2em}$1.04067\pm 0.00058$\hspace{2em} & \hspace{1em}$1.04004^{+0.00051}_{-0.00062}$\hspace{1em} & \hspace{1em}$1.04014\pm 0.00054$\hspace{1em} & \hspace{1em}$1.04007^{+0.00060}_{-0.00069}$\hspace{1em}  \\

$\Omega_b h^2$ & $0.02218\pm 0.00013$ & $0.02218\pm 0.00014$ & $0.02219\pm 0.00014$ & $0.02218\pm 0.00014$  \\

$\Omega_c h^2$ & $0.1219^{+0.0088}_{-0.011}$ & $0.133^{+0.010}_{-0.0086}$ & $0.1314\pm 0.0092$ & $0.133\pm 0.011$ \\

$\log(10^{10}A_s)$ & $3.043\pm 0.013$ & $3.045\pm 0.014$ & $3.045\pm 0.014$ & $3.045\pm 0.014$ \\

$n_s$ & $0.9664\pm 0.0037$ & $0.9669\pm 0.0037$ & $0.9668\pm 0.0037$ & $0.9669\pm 0.0037$ \\

$\tau_{\rm reio}$ & $0.0537\pm 0.0071$ & $0.0543\pm 0.0075$ & $0.0547\pm 0.0074$ & $0.0542\pm 0.0073$ \\

$w_0^{\rm dm}$ & $< 0.00115$ & $< -0.0436$ & $< -0.0359$ & $< -0.0401$ \\

$w_a^{\rm dm}$ & $0.00090^{+0.00037}_{-0.00043}$ & $0.00106^{+0.00043}_{-0.00052}$ & $0.00103^{+0.00038}_{-0.00051}$ & $0.00111^{+0.00044}_{-0.00055}$ \\

$a_t$ & $> 0.531$ & $0.50^{+0.11}_{-0.22}$ & $0.50^{+0.15}_{-0.23}$ & $0.48^{+0.13}_{-0.24}$ \\
\hline
\multicolumn{5}{|l|}{\textit{\textbf{Derived}}} \\

$H_0$ & $68.79^{+0.59}_{-0.51}$ & $68.16\pm 0.53$ & $68.25^{+0.58}_{-0.52}$ & $68.12^{+0.71}_{-0.60}$  \\

$\Omega_m$ & $0.306^{+0.020}_{-0.027}$ & $0.337^{+0.026}_{-0.022}$ & $0.331\pm 0.024$ & $0.337\pm 0.028$ \\

$\sigma_8$ & $0.802\pm 0.052$ & $0.745^{+0.033}_{-0.050}$ & $0.754^{+0.038}_{-0.049}$ & $0.746^{+0.041}_{-0.058}$ \\

$S_8$ & $0.808\pm 0.022$ & $0.787^{+0.016}_{-0.021}$ & $0.791^{+0.017}_{-0.021}$ & $0.788^{+0.017}_{-0.024}$ \\

$w_0$ & $-0.015^{+0.030}_{-0.046} $ & $-0.048^{+0.019}_{-0.033} $ & $-0.043^{+0.022}_{-0.034} $ & $-0.046^{+0.019}_{-0.037} $ \\

%$\chi^2$ &  &  &  &  \\

\bottomrule
\end{tabular}
\caption{The mean $\pm 1\sigma$ constraints on  cosmological parameters inferred from the addition of growth data to the data combinations considered in \cref{tab:dydn_const}.}
%\label{tab:dydn_g17}
\end{table*}
\begin{table*}
\centering
\scriptsize
%\footnotesize
%\small
\renewcommand{\arraystretch}{0.5}
\begin{tabular}{|lcccc|}
\hline
\hline
Parameter & \hspace{2em}PL-B\hspace{2em} & \hspace{2em}PL-B-DES\hspace{2em} & \hspace{2em}PL-B-PP\hspace{2em} & \hspace{2em}PL-B-U\hspace{2em} \\
\hline
\multicolumn{5}{|l|}{\textit{\textbf{Sampled}}} \\

$100\theta_{\ast}$ & \hspace{2em}$1.04100\pm 0.00023$\hspace{2em} & \hspace{1em}$1.04098\pm 0.00023$\hspace{1em} & \hspace{1em}$1.04098\pm 0.00023$\hspace{1em} & \hspace{1em}$1.04099\pm 0.00023$\hspace{1em}  \\

$\Omega_b h^2$ & $0.02234\pm 0.00012$ & $0.02232\pm 0.00012$ & $0.02232\pm 0.00012$ & $0.02232\pm 0.00012$  \\

$\Omega_c h^2$ & $0.11764\pm 0.00064$ & $0.11791\pm 0.00061$ & $0.11784\pm 0.00061$ & $0.11782\pm 0.00062$ \\

$\log(10^{10}A_s)$ & $3.053^{+0.013}_{-0.015}$ & $3.052\pm 0.014$ & $3.052\pm 0.014$ & $3.052\pm 0.014$ \\

$n_s$ & $0.9693\pm 0.0034$ & $0.9687\pm 0.0034$ & $0.9689\pm 0.0033$ & $0.9688\pm 0.0034$ \\

$\tau_{\rm reio}$ & $0.0604^{+0.0067}_{-0.0078}$ & $0.0597\pm 0.0073$ & $0.0600\pm 0.0074$ & $0.0601^{+0.0069}_{-0.0078}$ \\

\hline
\multicolumn{5}{|l|}{\textit{\textbf{Derived}}} \\

$H_0$ & $68.18\pm 0.29$ & $68.06\pm 0.27$ & $68.09\pm 0.27$ & $68.09\pm 0.28$  \\

$\Omega_m$ & $0.3026\pm 0.0037$ & $0.3042\pm 0.0035$ & $0.3037\pm 0.0035$ & $0.3037\pm 0.0036$ \\

$\sigma_8$ & $0.8081\pm 0.0058$ & $0.8085\pm 0.0057$ & $0.8085\pm 0.0057$ & $0.8084^{+0.0054}_{-0.0061}$ \\

$S_8$ & $0.8116\pm 0.0080$ & $0.8141\pm 0.0077$ & $0.8135\pm 0.0078$ & $0.8133\pm 0.0080$ \\

%$\chi^2$ &  &  &  &  \\

\bottomrule
\end{tabular}
\caption{The mean $\pm 1\sigma$ constraints on  cosmological parameters inferred from various datasets including DESI DR2, CMB, and supernovae and their combinations for the $\Lambda$CDM model.}
\label{tab:lcdm_const}
\end{table*}
%\\
%\vspace{0.1em}
\begin{table*}[h!]
\centering
\scriptsize
%\footnotesize
\renewcommand{\arraystretch}{0.5}
\begin{tabular}{|lcccc|}
\hline
\hline
Parameter & \hspace{2em}PL-BG\hspace{2em} & \hspace{2em}PL-BG-DES\hspace{2em} & \hspace{2em}PL-BG-PP\hspace{2em} & \hspace{2em}PL-BG-U\hspace{2em} \\
\hline
\multicolumn{5}{|l|}{\textit{\textbf{Sampled}}} \\

$100\theta_{\ast}$ & \hspace{2em}$1.04101\pm 0.00023$\hspace{2em} & \hspace{1em}$1.04099\pm 0.00023$\hspace{1em} & \hspace{1em}$1.04100\pm 0.00023$\hspace{1em} & \hspace{1em}$1.04100\pm 0.00023$\hspace{1em}  \\

$\Omega_b h^2$ & $0.02235\pm 0.00012$ & $0.02233\pm 0.00012$ & $0.02234\pm 0.00012$ & $0.02234\pm 0.00012$  \\

$\Omega_c h^2$ & $0.11743\pm 0.00062$ & $0.11770\pm 0.00061$ & $0.11764\pm 0.00061$ & $0.11762\pm 0.00061$ \\

$\log(10^{10}A_s)$ & $3.049\pm 0.014$ & $3.048\pm 0.014$ & $3.048\pm 0.013$ & $3.048\pm 0.014$ \\

$n_s$ & $0.9696\pm 0.0034$ & $0.9689\pm 0.0033$ & $0.9691\pm 0.0033$ & $0.9692\pm 0.0033$ \\

$\tau_{\rm reio}$ & $0.0585\pm 0.0072$ & $0.0578^{+0.0066}_{-0.0074}$ & $0.0578\pm 0.0071$ & $0.0580^{+0.0066}_{-0.0076}$ \\

\hline
\multicolumn{5}{|l|}{\textit{\textbf{Derived}}} \\

$H_0$ & $68.27\pm 0.28$ & $68.15\pm 0.28$ & $68.18\pm 0.28$ & $68.19\pm 0.28$  \\

$\Omega_m$ & $0.3013\pm 0.0036$ & $0.3029\pm 0.0035$ & $0.3026\pm 0.0036$ & $0.3024\pm 0.0036$ \\

$\sigma_8$ & $0.8058\pm 0.0055$ & $0.8061\pm 0.0055$ & $0.8060\pm 0.0054$ & $0.8061\pm 0.0055$ \\

$S_8$ & $0.8075\pm 0.0077$ & $0.8101\pm 0.0076$ & $0.8094\pm 0.0076$ & $0.8093\pm 0.0077$ \\

%$\chi^2$ &  &  &  &  \\

\bottomrule
\end{tabular}
\caption{The mean $\pm 1\sigma$ constraints on  cosmological parameters inferred from the addition of growth data to the data combinations considered in \cref{tab:lcdm_const}.}

\label{tab:lcdm_g17}

\end{table*}
\vspace{1em}

\begin{table*}[h!]
\centering
%\scriptsize
\footnotesize

\begin{tabular}{|lcccc|}
\hline
\hline
Parameter & \hspace{2em}PL-B\hspace{2em} & \hspace{2em}PL-B-DES\hspace{2em} & \hspace{2em}PL-B-PP\hspace{2em} & \hspace{2em}PL-B-U\hspace{2em} \\
\hline
\multicolumn{5}{|l|}{\textit{\textbf{Sampled}}} \\

$100\theta_{\ast}$ & \hspace{2em}$1.04078\pm 0.00024$\hspace{2em} & \hspace{1em}$1.04083\pm 0.00024$\hspace{1em} & \hspace{1em}$1.04085\pm 0.00024$\hspace{1em} & \hspace{1em}$1.04081\pm 0.00024$\hspace{1em}  \\

$\Omega_b h^2$ & $0.02221\pm 0.00013$ & $0.02223\pm 0.00012$ & $0.02224\pm 0.00012$ & $0.02223\pm 0.00013$  \\

$\Omega_c h^2$ & $0.11964\pm 0.00086$ & $0.11921\pm 0.00083$ & $0.11904\pm 0.00084$ & $0.11933\pm 0.00084$ \\

$\log(10^{10}A_s)$ & $3.039\pm 0.014$ & $3.042\pm 0.014$ & $3.044\pm 0.014$ & $3.042\pm 0.014$ \\

$n_s$ & $0.9642\pm 0.0037$ & $0.9653\pm 0.0036$ & $0.9658\pm 0.0036$ & $0.9651\pm 0.0036$ \\

$\tau_{\rm reio}$ & $0.0529\pm 0.0072$ & $0.0545\pm 0.0072$ & $0.0553\pm 0.0072$ & $0.0542\pm 0.0073$ \\

$w_{0,\mathrm{DE}}$ & $-0.41\pm 0.21$ & $-0.805\pm 0.056$ & $-0.838\pm 0.056$ & $-0.667\pm 0.089$ \\

$w_{a,\mathrm{DE}}$ & $-1.77\pm 0.59$ & $-0.74\pm 0.22$ & $-0.62\pm 0.21$ & $-1.09^{+0.31}_{-0.27}$ \\

\hline
\multicolumn{5}{|l|}{\textit{\textbf{Derived}}} \\

$H_0$ & $63.6^{+1.7}_{-2.1}$ & $67.33\pm 0.54$ & $67.51\pm 0.60$ & $65.92\pm 0.83$  \\

$\Omega_m$ & $0.353\pm 0.021$ & $0.3135\pm 0.0054$ & $0.3115\pm 0.0058$ & $0.3274\pm 0.0087$ \\

$\sigma_8$ & $0.781^{+0.015}_{-0.017}$ & $0.8110\pm 0.0083$ & $0.8116\pm 0.0087$ & $0.7999\pm 0.0097$ \\

$S_8$ & $0.847\pm 0.013$ & $0.8289\pm 0.0091$ & $0.8269\pm 0.0090$ & $0.8355\pm 0.0096$ \\

%$\chi^2$ &  &  &  &  \\

\bottomrule
\end{tabular}
\caption{The mean $\pm 1\sigma$ constraints on  cosmological parameters inferred from various datasets including DESI DR2, CMB, and supernovae and their combinations for the CPL parameterisation.}
\label{tab:cpl_const}
\end{table*}

\begin{table*}[h!]
\centering
%\scriptsize
\footnotesize

\begin{tabular}{|lcccc|}
\hline
\hline
Parameter & \hspace{2em}PL-BG\hspace{2em} & \hspace{2em}PL-BG-DES\hspace{2em} & \hspace{2em}PL-BG-PP\hspace{2em} & \hspace{2em}PL-BG-U\hspace{2em} \\
\hline
\multicolumn{5}{|l|}{\textit{\textbf{Sampled}}} \\

$100\theta_{\ast}$ & \hspace{2em}$1.04079\pm 0.00024$\hspace{2em} & \hspace{1em}$1.04086\pm 0.00024$\hspace{1em} & \hspace{1em}$1.04087\pm 0.00024$\hspace{1em} & \hspace{1em}$1.04083\pm 0.00024$\hspace{1em}  \\

$\Omega_b h^2$ & $0.02223\pm 0.00013$ & $0.02226\pm 0.00012$ & $0.02227\pm 0.00012$ & $0.02225\pm 0.00013$  \\

$\Omega_c h^2$ & $0.11931\pm 0.00086$ & $0.11882\pm 0.00082$ & $0.11867\pm 0.00082$ & $0.11898\pm 0.00082$ \\

$\log(10^{10}A_s)$ & $3.037\pm 0.014$ & $3.040\pm 0.014$ & $3.041\pm 0.014$ & $3.039\pm 0.014$ \\

$n_s$ & $0.9648\pm 0.0037$ & $0.9661\pm 0.0036$ & $0.9664\pm 0.0037$ & $0.9656\pm 0.0037$ \\

$\tau_{\rm reio}$ & $0.0520\pm 0.0072$ & $0.0535\pm 0.0072$ & $0.0540\pm 0.0072$ & $0.0528\pm 0.0072$ \\

$w_{0,\mathrm{DE}}$ & $-0.43\pm 0.20$ & $-0.805\pm 0.056$ & $-0.838\pm 0.054$ & $-0.669\pm 0.088$ \\

$w_{a,\mathrm{DE}}$ & $-1.70\pm 0.56$ & $-0.74 \pm 0.22$ & $-0.59^{+0.21}_{-0.19}$ & $-1.05^{+0.31}_{-0.28}$ \\

\hline
\multicolumn{5}{|l|}{\textit{\textbf{Derived}}} \\

$H_0$ & $63.7^{+1.6}_{-1.9}$ & $67.31\pm 0.54$ & $67.48\pm 0.59$ & $65.89\pm 0.83$  \\

$\Omega_m$ & $0.352\pm 0.020$ & $0.3129\pm 0.0053$ & $0.3110\pm 0.0056$ & $0.3269\pm 0.0085$ \\

$\sigma_8$ & $0.779\pm 0.016$ & $0.8069\pm 0.0083$ & $0.8074\pm 0.0085$ & $0.7958\pm 0.0094$ \\

$S_8$ & $0.842^{+0.013}_{-0.012}$ & $0.8240\pm 0.0088$ & $0.8220\pm 0.0085$ & $0.8307\pm 0.0093$ \\

%$\chi^2$ &  &  &  &  \\

\bottomrule
\end{tabular}
\caption{The mean $\pm 1\sigma$ constraints on  cosmological parameters inferred from the addition of growth data to the data combinations considered in \cref{tab:cpl_const}.}

\label{tab:cpl_g17}

\end{table*}
% Use the style of numbering in square brackets.
% If nothing is used, default style will be taken.
%\begin{enumerate}[a)]
%\item 
%\item 
%\item 
%\end{enumerate}  

%\begin{itemize}
%\item 
%\item 
%\item 
%\end{itemize}  

%\begin{description}
%\item[]
%\item[] 
%\item[] 
%\end{description}  

%\clearpage %%Remove this from your manuscript

% Uncomment and use as the case may be
%\begin{theorem} 
%\end{theorem}

% Uncomment and use as the case may be
%\begin{lemma} 
%\end{lemma}


%\section{}\label{}

\newpage
\printcredits

\bibliographystyle{cas-model2-names}


\bibliography{refs}



\end{document}